\documentclass[10pt]{IEEEtran}
\usepackage{url}
\usepackage{color}
\usepackage{cite}
\usepackage{tikz}
\usepackage{tikz-qtree}
\usepackage{xcolor}
\usepackage{listings}
\usepackage{graphicx}
\usepackage{tabularx}
\usepackage[utf8]{inputenc}
\usepackage{textpos}

\usepackage[justification=centering]{caption}

\title{XSS-FP: Browser Fingerprinting using HTML Parser Quirks}
\author{Abgrall Erwan, Yves Le Traon, Martin Monperrus, Sylvain Gombault, Mario Heiderich and Alain Ribault}
\date{}

\begin{document}
\maketitle

\begin{textblock*}{\textwidth}(0cm,-1.6cm)
\center\em\noindent Technical Report, University of Luxembourg, 2012.
\end{textblock*}

\begin{abstract}
There are many scenarios in which inferring the type of  a client browser is desirable, for instance to fight against session stealing.
This is known as browser fingerprinting. 
This paper presents and evaluates a novel fingerprinting 
technique to determine the exact nature (browser type and version, eg Firefox 15) of a web-browser, exploiting HTML parser quirks exercised through XSS.
Our experiments show that the exact version of a web browser can be determined with 71\% of 
accuracy, and that only 6 tests are sufficient to quickly determine the exact family a web 
browser belongs to. 
\end{abstract}
\section{Introduction}
\label{sec:intro}
In computer security, fingerprinting consists of identifying a system from the
outside, i.e. guessing its kind and version \cite{adhami2001fingerprinting}
 by observing 
specific behaviors (passive fingerprinting), or collecting system responses to
various stimuli (active fingerprinting). A common example of fingerprinting is
service fingerprinting. It consists of identifying the daemon behind an open
port of a server. For instance, a port scanner may output that the daemon 
behind port 22 is not the expected SSH server, but a SMTP daemon, instance of 
the software package ``Postfix'', in version 7.

OS fingerprinting is another popular kind of fingerprinting \cite{greenwald2007toward}. 
For instance, by sending carefully forged packets to the target, slight differences
between implementations of the TCP protocol stack enable observers to identify the
stack and its underlying operating system. 
Fingerprinting is used in many situations. 
For instance, security engineers use it to check whether known vulnerabilities may affect a software system
or infrastructure.

Similarly to service or OS fingerprinting, browser fingerprinting consists 
of identifying a browser implementation and version.
Also similarly to OS fingerprinting, there are two kinds of browser fingerprinting.
On the one hand, one may uniquely identify a browser (see e.g. \cite{eckersley2010unique}), 
on the other hand, one may uniquely identify a browser type, that is, identifying the browser implementation (e.g. Firefox vs Internet Explorer) and its version number (e.g. IE8 vs IE9).
They are orthogonal concerns: the former is important w.r.t. privacy, the second is important w.r.t. security, and there is no direct relation between both.

In this paper, we address the latter, the fingerprinting of browser type and version.
There are many use cases of browser fingerprinting (see Section \ref{sec:rationale}) for instance to address the problem of credentials stealing detection.
Previous work in the field of browser fingerprinting was based on analyzing the 
JavaScript behavior \cite{mowery2011fingerprinting} or  the network behavior~\cite{yen2009browser} of a browser.
In this paper, we propose to use the behavior of the HTML parser under specific inputs to 
fingerprint the type and version of browsers. 
We call those particular responses \emph{HTML parser quirks\footnote{The Merriam-Webster
dictionary defines a quirk as a ``a peculiar trait''}}.
Those specific inputs are the same that are used for cross-site scripting 
attacks.
This is completely novel browser fingerprinting technique, with key advantages:
1) compared to network-based fingerprinting, it can be achieved at the application level with 
no access to the low level network stack.
2) it is hardly spoofable; simulating the behavior of an HTML parser is difficult without running the parser itself.
We will give more details on these points in Section \ref{sec:rationales}.
To the best of our knowledge, we are the first to use HTML parser quirks to achieve browser fingerprinting.

Our experiments show that the exact version
of a web browser out of 77 can be determined thanks to its signature. 
Moreover, using classification techniques described by Hall et al.~\cite{hall2009weka},  only $6$ XSS tests are 
sufficient to determine the exact family a web browser belongs to.

Section~\ref{sec:rationales} further discusses the rationales of browser fingerprinting
and using HTML parser quirks.
Section \ref{sec:overview} is an overview of the approach.
The next sections describe the XSS vector 
collection, and the dedicated tool we have developed to execute the HTML parser quirks. 
Section \ref{sec:fingerprinting-methodology} describes data mining classification we use
to fingerprint browsers. 
Section \ref{sec:analysis} discusses our fingerprinting capabilities, including a discussion on how fingerprints can be forged.
Section \ref{sec:other-uses} discusses browser fingerprinting from an expert security engineer viewpoint.
Section \ref{sec:related} is a comparison against the related work, section \ref{sec:conclusion} concludes the paper.

\section{Rationales}
\label{sec:rationales}
The HTTP protocol specifies that browsers should send a specific string, 
value of the HTTP request header \emph{User-Agent}  (UA), for identifying themselves.
In practice, all browsers do send this header. 
The rationale behind this header is to offer the server a way to infer
browser capabilities and serve specific contents to more or less advanced 
browsers.

Can a server trust the \emph{User-Agent} header to fingerprint a browser ?
No, this value is set by the browser and it cannot be trusted since an
attacker can modify it by patching the browser (some browsers even offer to set
it as a configuration point).
The User Agent string is commonly used by exploit kits to attack servers by embedding a malicious payload in the 
 user-agent header.

\begin{figure*}
\includegraphics[width=\textwidth]{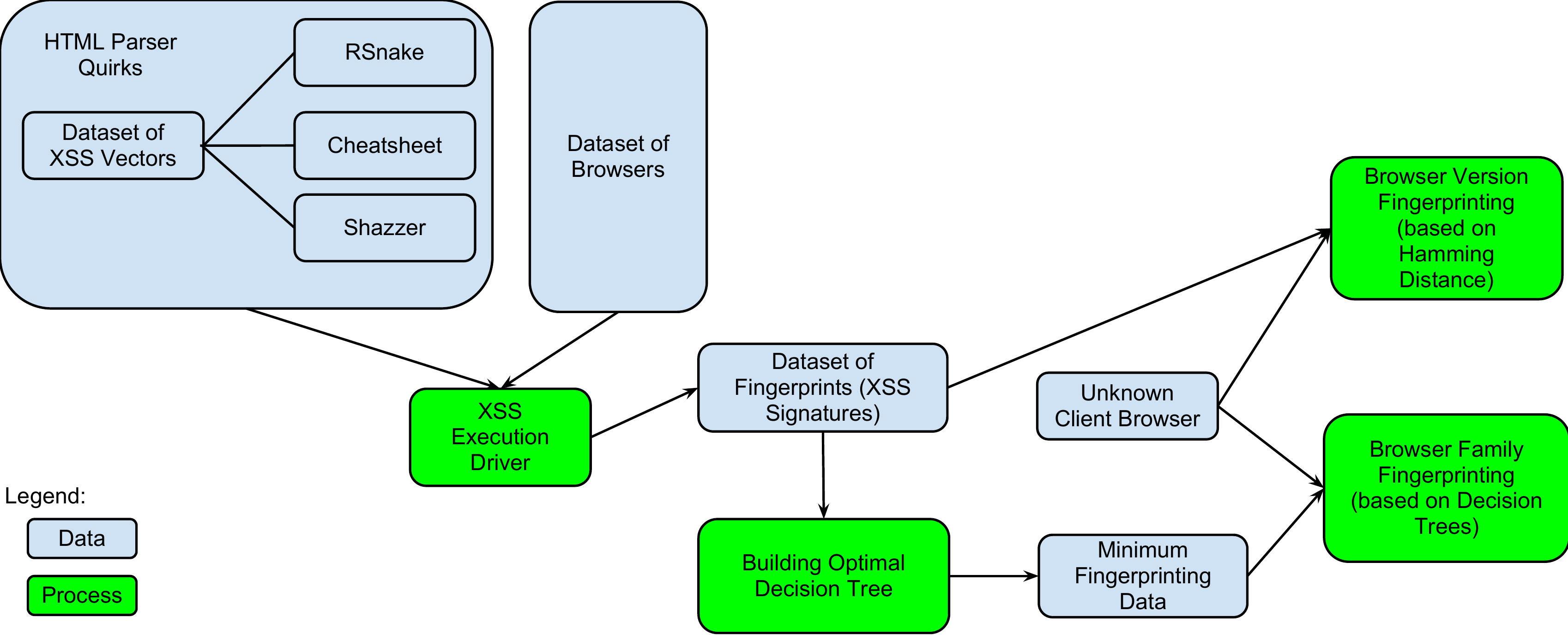}
\caption{Overview of Our browser Fingerprinting Process}
\label{fig:overview}
\end{figure*}

\subsection{Defeating Session Stealing with Browser Fingerprinting}
Session stealing means stealing a cookie or a session ID in order to access unauthorized resources.
Server-side software is responsible to detect session stealing.
This can be done through checking whether the presented cookie or session ID matches the HTTP 
user-agent header. However, as said, this does not work if attackers are able to steal both the 
credentials and the user-agent. Checking credentials with IP addresses is not a valid way to 
check session stealing due to users mobility and NAT mechanisms.

With browser fingerprinting, at any point in time, server software can:
1) verify whether the HTTP user-agent matches the inferred browser type (detection of UA spoofing)
2) verify whether the inferred browser type matches the browser that was used on login (detection of session stealing).

Beyond this key use-case, there are many other uses of browser fingerprinting, further discussed
in Section \ref{sec:other-uses}.

\subsection{The Benefits of Using HTML Parser Quirks For Fingerprinting}
Previous work in the field of browser fingerprinting was based on analyzing the 
JavaScript  behavior \cite{mowery2011fingerprinting} or the network behavior~\cite{yen2009browser} of browsers.
In this paper, we use the HTML parser quirks for browser fingerprinting.
HTML parser quirks are peculiar behaviors under specific inputs.
They may have different consequences, in particular incorrect rendering or 
undesired execution of JavaScript code.

The latter point is daily exploited by 
cross-site scripting attacks (XSS).
A cross-site scripting attack embeds an executable malicious payload into a piece 
of specific HTML code.
By replacing the malicious payload by a simple binary output telling the server 
whether a specific parser behavior is observed or not, one can observe from the 
server-side the execution of HTML parser quirks. 
For us, those execution-based quirks are invaluable: they are testable.

Furthermore, HTML parser quirks are known. The very active community on cross-site 
scripting research has produced inventories of HTML parser quirks.
This means we have tons of HTML parser quirks to achieve browser fingerprinting.

One might think that what we call ``quirks'' are essentially ``bugs''. We think that this distinction
is not binary. Indeed, the root cause of some known XSS vectors can be found in the specification
itself (e.g. the HTML5 specification), that is it is not a standard  implementation bug.
Consequently, we consider that the browser behavior under particular input  is a ``quirk'', whether desired or not, and whether incorrect or not.

Compared to network-based fingerprinting, HTML-based fingerprinting can be achieved 
at the application level with no access to the low level network stack. This means 
that an application can use browser fingerprinting 
(for instance for detecting session stealing), while remaining OS independent. 
For instance, a server-side application written can still perform browser 
fingerprinting independently of the application server (Tomcat, JBoss, etc.), 
the Java virtual machine (IBM J9, OpenJDK, etc.) and the OS (Windows, Linux, etc.).

Last but not least, the behavior of an HTML parser is very complex (that's why so 
many cross-site scripting attacks exist). 
Hence, the fingerprint of HTML parser quirks is hard to spoof. 
In other terms, if an attacker wants to deploy counter-measures to an HTML-based 
fingerprinting, he has no solution but running all browsers in parallel.

The implementation of such  HTML parser checks can be achieved through the inclusion 
of a small invisible \emph{iframe}. 
Checks can be triggered upon sensitive actions or randomly.
We also  imagine web application firewalls modifying some pages on the fly to add 
the signature checks based on HTML parser quirks.

\subsection{Recapitulation}
There is a need for browser fingerprinting, since the HTTP protocol has no means to 
fight against session stealing.
A technique based on the observation of HTML parser quirks is doable at the 
application level, and its counter-attack is hard, 
since HTML parser behavior is hardly spoofable.

\section{Overview of the Approach}
\label{sec:overview}
Figure \ref{fig:overview} presents an overview of our browser fingerprinting approach.
Using quirks to fingerprint web browsers is feasible only if these
quirks are testable, in the sense that the specific behavior of the
browser quirk can be observed through testing. 
This is why we build our own \emph{dataset of testable quirks}. They come from different sources: collaborative,
fuzzing techniques such as Shazzer, existing referenced vectors 
(see section \ref{subsec:rsnake_xss_cheat_sheet}).

Based on this set of testable XSS vectors, a framework called \emph{XSS Test Driver} 
performs the full test suite on different browsers, collecting as many XSS signatures as possible. 
Each signature contains attributes describing the results of all the tests. 
We consider \emph{an initial set of 77 browsers}, and the corresponding signatures are  
referred as \emph{the raw dataset of browser signatures}.
This dataset can be directly used for fingerprinting an unknown
web browser, in order to determine (1) its exact version based on a
Hamming distance between browser signatures. This set can also be used  (2) as input 
for machine learning techniques in order to
\emph{build an optimized decision tree}. Such a decision tree allows the quick classification of the family 
(e.g. Firefox or Chrome) of 
an unknown web browser according to its responses to minimum
fingerprinting data (execution of a handful of quirks instead of thousands). 
It has to be noted that the overall approach can be applied using any testable
quirks. All the fingerprinting process steps are described in the
following sections.

\section{A Dataset of HTML Parser Quirks}
\label{sec:collecting-xss-vectors}
The following subsection describes the three sources we have used to build a significantly large collection of XSS vectors usable for fingerprinting. 
These sources include static vector 
libraries as well as XSS fuzz generation tools.

\subsection{RSnake's XSS Cheat Sheet - Legacy Vector Collection}
\label{subsec:rsnake_xss_cheat_sheet}
The XSS Cheat Sheet was created by R. `RSnake'' Hansen et al., and provides a richresource for penetration testers and developers. It showcases an overall of \~100 different XSS vectors demonstrating character and string parsing issues, especially for legacy browsers. The resource has not been updated for many years though; modern HTML5 and SVG based attack vector examples are not present in this document. A beta-version of an overworked XSS Cheat Sheet was announced in 2010, but never found its way to a public release. The lack of updates of this document lead to community-driven projects such the HTML5 Security Cheatsheet (H5SC).

\subsection{HTML5 Security Cheatsheet - Community Vector Collection}
\label{subsec:html5_security_cheatsheet}
The HTML5 Security Cheatsheet (H5SC) is a community driven project that aims at documenting and categorizing known XSS and other client-side attack vectors. 
The H5SC provides a simple JSON based storage model and allows registered and
approved contributors to add new XSS vectors, modify existing data and most 
importantly provide version information on which user agents are affected by 
the demonstrated attack vector. This allows security professionals and
developers to protect their applications accordingly and even perform basic 
risk assessment, for instance when fixing a vector is in conflict with required application 
features. The H5SC set contains  \~120 individual attack vectors 
alongside with detailed explanations on their inner workings.

\subsection{Shazzer - Collabrative Fuzzing for Identifying XSS Vectors}
\label{subsec:shazzer_community_fuzzing_for_browser_bugs_leveraging_xss}
Shazzer\footnote{see \url{http://shazzer.co.uk/home}} is a collaborative website aiming at providing 
an interface for collaboratively specifying and  identifying possible XSS vectors. 
Shazzer offers  enumeration templates and an internal render and storage engine.
A user can for instance define a vector template containing various different placeholders. 
After starting the actual fuzzing process, the placeholders will be iteratively replaced
by the corresponding characters and rendered in an isolated \emph{iframe} to see whether the desired effect 
can be accomplished with the currently tested characters. Shazzer has been used by a large number of security
testers to determine whether certain known an unknown parser bugs in modern user agents have been discovered and fixed.

The set of sources of XSS vectors is summarized Table \ref{tab:vectorsource}. For a total of 523 vectors, 
the main provider is Shazzer (291). The full vector list is available 
at \url{http://xss2.technomancie.net/vectors/}

\begin{table}
  \centering
   \caption{Composition of the XSS database (number of XSS vectors per source)}
    \begin{tabular}{c|c|c|c}
    \hline
     Rsnake & Html5Sec & Shazzer & Total \\
    \hline
     69    & 163   & 291 & 523 \\
    \hline
    \end{tabular}
  \label{tab:vectorsource}
\end{table}

\section{XSS Execution Driver}
\label{sec:xss-vector-testing}
In this section we present \emph{XSS Test Driver}, our framework to automatically perform 
the execution of XSS vectors for fingerprinting. We use our whole XSS vector set on a
 set of browsers, building a \emph{dataset of raw browser signatures}. 
 
\subsection{Terminology}
An XSS attack consists of executing code (mostly JavaScript) inside a browser via a website, by injecting a content (e.g. by posting a comment on a page).
The injected content is an \emph{XSS vector}.
For instance a very simple XSS vector is \verb*|<script>alert('foo');</script>|.
An XSS vector can be logically decomposed of three parts:
\begin{enumerate}
\item The \emph{XSS vector} contains one or several HTML tags and attributes
\item The \emph{payload} is a piece of JavaScript code,
\item The \emph{payload format} is a special way to encode the payload.
\end{enumerate}

In the above example, the vector is composed of the \textit{script} tags,
the payload is a call to function alert , and the format is ``identity'' 
(i.e. the payload in not encoded at all).

This is a very simple example of XSS vector.  More complex XSS Vectors benefit from the ever-growing functionalities offered by browsers to developers. 
Each new API or language subset that is able to execute or call JavaScript code can be turned into an 
XSS vector. For more information on the richness of XSS Vector forms, refer to section \ref{sec:xss-kind} 
and look at the XSS Vector sources described in \ref{sec:collecting-xss-vectors}

An important characteristic of XSS vectors is that certain XSS structures accept 
payloads in very specific formats. For instance, some XSS structures require a link 
to JavaScript file, others are successful only if the payload is encoded in base64. 
Such behavior is either related to a specific feature, or to a bug.

An XSS vector can also depend on :
\begin{itemize}
\item The  \emph{character set} the browser should use to decode the HTMK 
\item The  \emph{content type} of the transmitted resource
\item The  \emph{HTML Doctype} of the HTML Document
\end{itemize}

Since the browsers rely on those pieces of information to decode the received data and to 
parse them properly, some quirks can be triggered by playing with those 
parameters on the server side (ie: sending HTML4 vectors within an HTML5 context).

\subsection{XSS Test Environment}
The test environment of an XSS vector consists of two parts:
the HTML context and the encoding. The HTML context (that we call ``Web Context'') 
is composed of  the doctype and generally all the HTML surrounding the vector as well 
as the  MIME type specified in the HTTP headers. The encoding is the character set 
declared in the http headers and used in the document.

XSS vectors can be tested in different web contexts and with different encodings. 
Hence, each XSS vector must be executed by the product of the number of contexts and the 
number of encodings. As we have $523$ vectors, this may yield a combinatorial explosion if we run all possible encodings and web contexts. 
In the following experiments we limit the test set to two web contexts: \textit{quirks} and html5, and to one encoding \textit{utf-8}. Thus testing $523$ Vectors 
with $2$ web context and $1$ encoding generated $1046$ test cases to run.
Increasing the number of encodings tested allows using more discriminating quirks
but it is a trade off to be made during the signature collection, since it increases the number of tests to execute.

\subsection{Test Logic}
\label{sec:test-logic}
We use the following logic to chain the tests and collect the results:
\begin{enumerate}
    \item Each XSS attack is served by a URL containing a JavaScript payload encoded with the proper format. 
     When the URL is requested, the test is marked as \emph{SENT}.
    \item The payload of an XSS attack contains a specific JavaScript validation routine (described in \ref{magic-payload}). 
  When the validation mechanism is triggered (the validation routine is executed), the successful test is marked as \emph{PASS}.
    \item The server then points to a new test by redirecting the browser using a HTTP code 302 redirect.
  \item Upon completion of the test suite, \emph{SENT} tests cases are considered failed and remains with this value.
  \item If for some reason a test is skipped, or if a new untestded vector is introduced, the test result is marked Not Available (\textit{NA}).
\end{enumerate}
This test logic avoids the use of JavaScript library, and uses no interactions with the DOM.
It can be fully automated by using a runner script opening the next test inside an \emph{iframe}.
Chaining test execution can also be done manually by browsing different tests.

\subsection{Callback Functions : Validating JavaScript Execution}
\label{magic-payload}
        
    Depending on the browser JavaScript Engine, and how and where in the DOM the 
  JavaScript call is done, some function might not work. The first method in 
  XSS Test Driver generates a JavaScript redirect of the web page to the 
  test validation URL. But with some vectors, this method doesn't trigger the 
  expected web page redirection. it is due to some \emph{iframe} sandbox mechanisms 
  where the JavaScript code can't access window.location DOM property.
    A cookie based execution validation was added then, adding a cookie in the 
  browser to validate execution of a given test case, but it triggered security 
  errors on Chrome \emph{iframe} sandbox with \textit{srcdoc} based vectors.
    A \textit{XMLHttpRequest} call is also present in the test payload, triggering 
  a specific validation URL. But this one too was subject to some security 
  restrictions with recent versions of Chrome. 
    We eventually added a \texttt{<img>}-based callback to the payload, adding an 
  image to the DOM with an image source set to a validation URL delivering a 
  green \emph{PASS} verdict image.
    
\begin{table}\center
\caption{Examples of results of the XSS Test Driver\label{rawinstance}}
\begin{tabular}{l|l|c|c|c}
\hline
Attr. & browser  & 1-1-1 &  1-2-1 & \ldots523-2-1 \\
\hline
Value & Safari 5\_1\_5 & NA &  PASS  & NA  \\
\cline{2-5}
         & Firefox 11\_0 &  PASS & SENT  & NA  \\      
\hline
\end{tabular}
\label{tab:instance-example}
\end{table}
\subsection{Browser Instance Description}
For each tested browser, XSS Test Driver provides 1 signature instance (set of attributes) 
describing the results for the whole test suite 
representing 1046 unitary test cases computed from the 523 base XSS vectors. 
Each attribute name issued from 1 test has the same name
structure giving as many different attribute names (like 90-2-1 for example) :
\begin{itemize}
        \item XSS vector number of our test bed: 1 to 523,
        \item context of execution: 1 or 2,
        \item context of encoding: 1.
\end{itemize}
The possible values of these attributes are : $\{SENT,PASS,NA\}$ corresponding to the test logic,  
Section \ref{sec:test-logic}.
This set of attributes is completed with a free text describing the browser. 
Table \ref{tab:instance-example} illustrates 2 instances extracted from the real dataset.

\section{Fingerprinting Methodology}
\label{sec:fingerprinting-methodology}
This section presents the methodology we use to fingerprint browsers using their responses 
 when executing XSS vectors based tests. The signature dataset provided by XSS Test Driver is used as input.

\subsection{Exact Fingerprinting Based on Hamming Distance between Browser Signatures}
\label{sec:method-exact-fingerprinting}

Similarity measurement is used to find nearest neighbors in a set of vectors.
An efficient way of doing it is to calculate the Hamming Distance between vectors. 
The Hamming Distance evaluates similarities between 2 vectors having the same number of dimensions, and is defined as follow:
for two vectors V1 and V2, this measure corresponds to the number of dimensions where the element of the vector V1 differs from the element of the vector V2.

\subsubsection{XSS Browser Signature}

We define the \emph{browser signature} as a vector computed from a browser instance
provided by XSS Test Driver.
The size of the vector is $n$  where $n$ is the number of XSS vectors in the database 
(see \ref{sec:collecting-xss-vectors}).
As defined in the test logic section \ref{sec:test-logic}, the value of each element is in the set: $\{s,p,n\}$ where
\begin{itemize}
\item \emph{s} corresponds to \emph{SENT}
\item \emph{p} corresponds to \emph{PASS}
\item \emph{n} corresponds to \emph{NA}
\end{itemize}

Let us consider the following simple signatures \emph{Sb1}, \emph{Sb2} and \emph{Sb3} that are obtained from executing three XSS vectors on three web browsers \emph{b1}, \emph{b2} and \emph{b3}.
\begin{itemize}
\item \emph{Sb1} = \emph{pps}
\item \emph{Sb2} = \emph{pns}
\item \emph{Sb3} = \emph{pnp}
\end{itemize}

\emph{Sb1} captures the fact that the two first XSS-vector executions are \emph{PASS} and the last  \emph{SENT}.

To deal with browsers for which we do not have enough significant data for fingerprinting, 
we define a \emph{confidence} value based on the percentage of XSS vectors the web browser executes: $\sum(PASS|SENT)/\sum(XSS vectors)$. 
If this value is too 
low for a given browser, we cannot trust its instance. Browsers with signature 
confidence above 90\% are used in this paper.

\subsubsection{Modified Hamming Distance}

To measure similarity between two browser signatures, we propose a modified Hamming distance (MHD) 
in order to ignore \emph{NA} in the signature.

Our distance works as follow: given two browser signatures, it computes the Hamming distance only on XSS results that are \emph{s} or \emph{p} in both signatures (not \emph{n}).
The modified Hamming distance between \emph{Sb1} and \emph{Sb2} is 0, and the MHD between \emph{Sb1} and \emph{Sb3} is 1.

When XSS Test Driver collects a signature,  we compute the MHD between this signature and the known signatures from the browser dataset. When two browser signatures in the database have a MHD of 0, the fingerprint cannot distinguish among those corresponding browsers: they are similar,
meaning that we may have many signatures of very close versions e.g., \emph{Firefox 10.1.1} and \emph{Firefox 10.1.2}. If there is no browser signature in  the database with a distance of 0, we consider the \emph{nearest neighbor} defined the browser signature(s) with the smallest MHD. Having a a nearest neighbor with a large MHD means that the browser is clearly distinguished among the dataset.

As a complement and to evaluate how the browsers belonging to a family are grouped, we calculate the \emph{Median Distance to the Family} (MDF) of the browser. As its name suggests, MDF indicates whether a browser is close to its siblings or not.
If all browsers of a family have a low MDF, it means that the family is cohesive and they do share the same HTML parser behavior.
We also compute the  \emph{Median Distance to the Dataset} (MDD) to determine 
\textit{outlier} browsers, those that do not resemble any other browser.

\subsection{Browser Family Fingerprinting using Decision Trees}
Identifying the browser family with a minimum of tests is crucial when an attacker
spoofs the UA string, in order to minimize the spoof detection time. 
The raw dataset produced by XSS Test Driver is also used to validate
the fingerprinting methodology based on machine learning algorithms.

\subsubsection{Classification based on Decision Trees}
The classification algorithms based on decision trees (DT) are useful in supervised data mining since they obtain reasonable accuracy
and are relatively inexpensive to compute. DT classifiers are based on the
\emph{divide and conquer} strategy to construct an appropriate tree from a given learning
set containing a set of labeled instances, whose characteristic is to have a class
attribute. As a well known and widely used algorithm, C4.5 (developed by Quinlan \cite{quinlan1993c4})
generates accurate decision trees that can be used for effective classification. We have
used J48 decision tree algorithm, a Weka \cite{hall2009weka} implementation of C4.5. 
It builds a decision tree from a set of training data also with the concept of information
entropy. It uses the fact that each attribute of the data can be used to make a decision
that splits the data into smaller subsets. Like C4.5, J48 examines the information gain
ratio (can be regarded as normalized information gain) that results from choosing an
attribute for splitting the data.

The attribute with the highest information gain ratio
is the one used to make the decision. The decision trees are constructed as a set of rules
during learning phase. Rules can be seen as a tree composed of nodes containing tests on
attributes and leading to leaves containing the class of the learned instance. It is then
used to predict the class of new instances belonging to a testing set, based on the rules.

\subsubsection{Labeled Browser Instance Description}
XSS Test Driver provides the \emph{initial dataset} needed to fingerprint the browser family. 
The chosen browser families correspond to recent browsers: \emph{Android, Chrome, Firefox, 
Internet Explorer(IE), Opera and Safari}.
Table \ref{family} summarizes the number of tested browsers per browser family, a subset of 72 
instances. 
To build the labeled dataset, we consider as attributes for classification the P, S and N values of the XSS test execution, and we add an attribute labeled \emph{family}.
 This  \emph{family} attribute may have one of the 6 possible values listed in table \ref{family}.

Table \ref{labeledinstance} presents 2 labeled instances extracted from the real data set.
 
\begin{table}
\center
\caption{Distribution of Browser Families\label{family}}
\begin{tabular}{c|r}
\hline
Family & Instances \\
\hline
Android & 15  \\
\hline
Chrome & 19  \\
\hline
Firefox & 15 \\
\hline
IE (Internet Explorer)  & 6   \\
\hline
Opera & 6   \\
\hline
Safari & 15  \\
\hline
\end{tabular}
\end{table}
           
\begin{table}\center
\caption{Example of labeled signatures\label{labeledinstance}}
\begin{tabular}{l|c|c|c|l}
\hline
Attr. &  1-1-1 &  1-2-1 & \ldots523-2-1 & family \\
\hline
Value &  N &  P & N & Safari \\
\cline{2-5}
         &  P & S & N & Firefox \\      
\hline
\end{tabular}
\end{table}

\subsubsection{Building the Decision Tree}
We configure Weka Explorer to use J48 classification algorithm and \emph{family} as class 
attribute. Firstly, We consider the whole labeled data set containing 72 instances to train J48 
classifier. The generated DT is composed of nodes containing tests on attributes values, 
until the leaf containing the class attribute filled during the learning phase. After the 
training phase, we use the same data set to test our DT and we compare the class obtained 
with the DT to the class present in the instance: a difference reveals a misclassification. 
The quantity of errors of this first evaluation gives an estimation of the classifier 
produced by the whole data set regarding the class attribute \textit{family}.

\section{Experimental Results}
\label{sec:experimental-results}
In this section we analyze the results of our browser fingerprinting experiments.

\subsection{Exact Fingerprinting Results}
We have applied the method described in \ref{sec:method-exact-fingerprinting} to fingerprint 
our dataset of browsers in order to see whether the resulting fingerprints are discriminant.
Tables \ref{tab:distancetable1} and \ref{tab:distancetable2} (at the end of the paper for sake 
of readability) present our results.
The first column lists all browsers of our dataset. 
The second column indicates the nearest neighbor within the dataset according to the 
Hamming distance between browser signatures.
The third column gives the distance between those two neighbors.
The fourth and fifth columns are the median distances to the browsers of the same family (MDF) 
and the number of elements in the family. The last 
column is the median distance to the whole dataset (to see whether they are family or true outliers).
The results are ordered by MDF.

First, one sees that  for all browsers with a MHD of 0 to their nearest neighbor, the neighbor is 
a browser of the same family with a very close version number. 
\emph{This confirms the soundness of our approach.}

Second, i.e. 78\% of our browser dataset have a nearest neighbor 
at a MHD distance higher than 0.
This means that those browsers can perfectly be discriminated and that 
MHD is an appropriate distance to capture both the family and the version 
information.
\emph{This confirms our intuition that browser fingerprinting using XSS vectors is very 
discriminant}.

Interestingly, browsers 89, 25 and 27 are \emph{exotic} browsers like the ones you can find inside set top boxes
or smart-tv. 
Their MDD is very high, showing that MDD actually captures the originality of browser 
implementation.
For instance, browsers with an older code base like Konqueror are at a huge distance from the dataset mainly 
composed of recent browsers.
Also, the nearest neighbor of Rekonq is Safari 5.0.6, which makes sense since they use the same major 
version of the webkit engine (534). 
The MDFs of each browser in the dataset indicates its proximity with the rest of its family. 
The Firefox family and the Chrome family both contains a bigger number of elements due to the 
higher pace of release. Time and differences between two major versions of Firefox or Chrome 
is equivalent to minor version changes for IE or Safari in term of release time line.

Rekonq Linux, Origin  and Konqueror browsers use Webkit as HTML parser, 
as also do Safari, 
Chrome and Android.
We can see that browsers in the same family have similar MDF (e.g. $MDF=15$ for Firefox).
\emph{This shows that MDF correctly captures clusters of related browsers}.

The summary of this experiment is that if two browsers share the same HTML parsing code base, they 
also share highly similar fingerprints.
    
\begin{table}
  \centering
  \caption{MHD Fingerprinting Efficiency analysis}
    \begin{tabular}{r|r|r|r}
    \hline
    MHD=0 & nb of browsers & FP rate & Well Fingerprinted \\
    \hline
    22    & 77    & 28,57\% & 71,42\% \\
    \hline
    \end{tabular}
  \label{tab:mhdefficiency}
\end{table}
\begin{table}\center
\caption{Browser family classification results\label{results}}
\begin{tabular}{p{5.8cm}|r|r}
\hline
Total number of instances & 72 & 100.00\% \\
\hline
Correctly classified instances & 71 & 98.61\%  \\
\hline
Incorrectly classified instances & 1 & 1.38\%  \\
\hline
\end{tabular}
\end{table}

\subsection{Browser Family Fingerprinting Results}

\begin{figure*}
\begin{tikzpicture}[grow'=right,level distance=.65in]
\Tree [.397-1-1 [.PASS Firefox ][.!=PASS [.89-1-1 [.!=PASS [.90-2-1 [.!=PASS Chrome ][.PASS [.128-1-1 [.!=PASS Android ][.PASS [.258-1-1 [.SENT Safari ][.!=SENT Opera ]]]]]]][.PASS IE ]]]]
\end{tikzpicture}
\caption{Executing only 6 XSS vectors enables us to classify the browser family with 98\% precision.}
\label{tree}
\end{figure*}
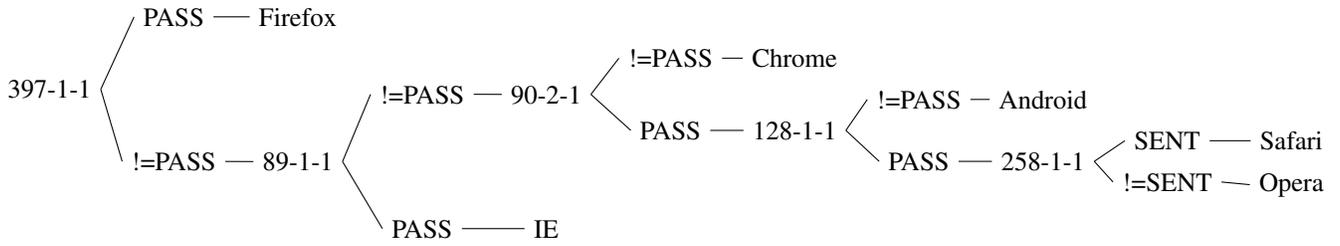
We use the whole dataset to train and build the decision tree presented of Figure \ref{tree}. We use this
tree to classify the training set, giving the results presented in table \ref{results}. 
\emph{The key point of this decision tree is that one can classify 98\% of the dataset using 
only 6 runs of XSS vectors.}

The confusion matrix highlights the accuracy of the classification using our DT. The diagonal of 
the matrix counts how many instances belonging to a class are correctly classified in this 
class. One can observe that the instance incorrectly classified belongs to \textit{Android}
and is classified as \textit{Chrome}. Since Chrome and Android share a significant code base, it is 
logic that some instances of Android are close to some Chrome instances.

Vectors \#89, \#90, \#128 and \#258 come from Shazzer and use parser bugs to special characters like 0x00. 
Vector \#397 is specific to Gecko-based browsers and come from html5sec\footnote{\url{http://html5sec.org/#15}}.

As a first experimentation, we plan to develop this approach as a piece of software in a web 
application firewall. This first step needs further investigations to validate our decision
tree on a larger set of browsers.

\subsection{Recapitulation}
Our experiments show that the exact version of a web browser can be determined with 71\% of 
accuracy (within our dataset), and that only 6 tests are sufficient to quickly determine the exact family a web 
browser belongs to.

\section{Discussion}
\label{sec:analysis}
\subsection{On Time and XSS}
The fact that one can determine the browser exact version just using
quirks is appealing. In particular, one can wonder whether there is
some underlying logic in the way the quirks occur, making them
predictable. Indeed, we could expect two successive versions of a
given browser to exhibit more similar quirks than more temporally
distant ones. 
There may be general temporal
factors explaining the discrimination power of HTML parser quirks. One of such
explanation factor could be the evolution of JavaScript and HTML norms
over time.

In this section, we investigate two research questions
to better analyze the discrimination power of quirks (at least those
provoked by our XSS vector dataset) on which we build the
fingerprinting technique.
The two research questions are:
RQ1) Can we observe general trends relating the temporal distance of
two web browser instances with their exhibited quirks?
RQ2) Does the discrimination power of quirks decrease when the
versions of a given web browser family are close?

\begin{figure*}
\includegraphics[width=\textwidth]{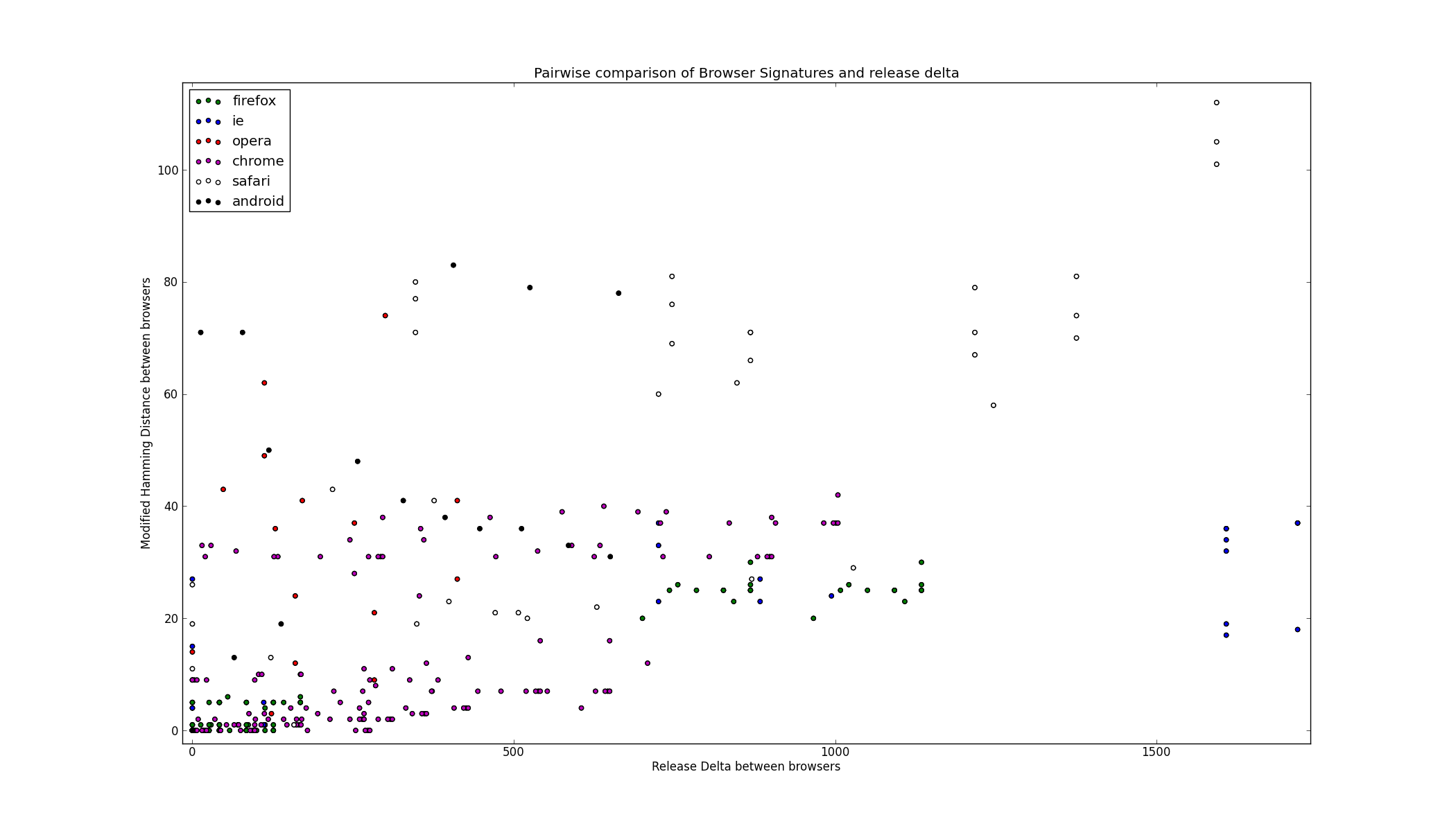}
\caption{Analysis of the relation between browser birth date and modified Hamminng distance.}
\label{fig:timemap}
\end{figure*}

Figure \ref{fig:timemap} answers to those questions.
Each plot represents a pair of web browser
instances. The X-axis value is the time period in days of the release dates of the two browsers.
The Y-axis value is the Hamming distance between both as defined above. 
What we see in this figure is that there is no general rule of the form, the longer between two versions, 
the more differences between HTML quirks.
Also, HTML quirks cannot be only related to JavaScript or HTML evolution.

This is a strong argument in favor of our approach because it means that one can trust the fingerprinting prediction, 
even if the client browser is of an unknown type.

\begin{figure*}
\includegraphics[width=\textwidth]{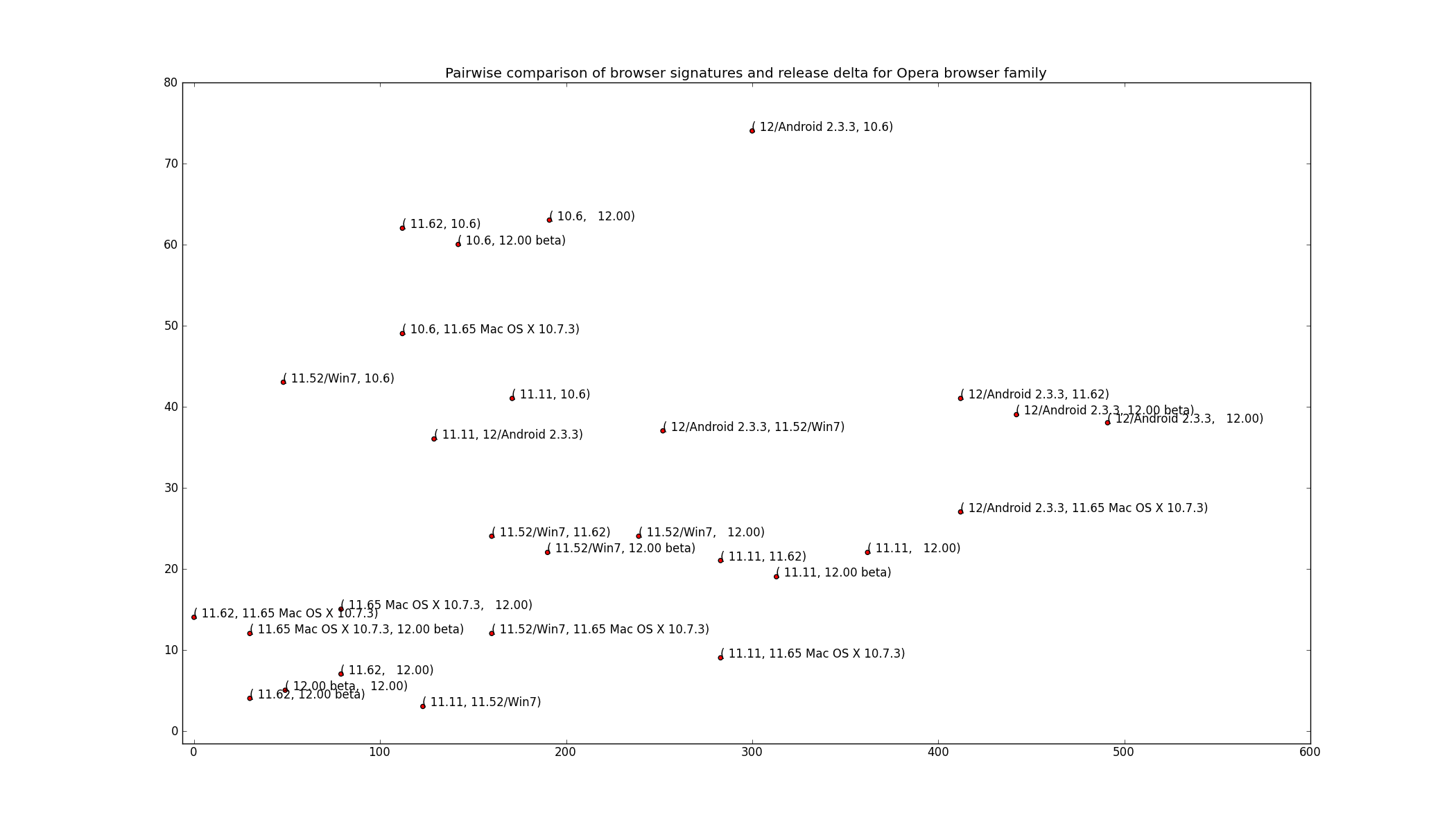}
\caption{Analysis of the relation between browser birth date and modified Hamminng distance for the Opera family}
\label{fig:timemap_opera}
\end{figure*}

Concerning RQ2, we go more in depth in the analysis and consider local
factors, that may be related to the development process into a same
web browser family. Usually regression tests are run to ensure that a
new version does not behave in a different manner than the previous
one, at least for its existing functionalities. We should thus observe
that two versions close from a temporal viewpoint have nearly the same
Hamming distance. As an example, Figure \ref{fig:timemap_opera} (Opera alone) plots every
pair of web browser versions for Opera. Surprisingly, no clear trend
appears. 
This also applies to other browser families.
It seems that there is no systematic development processes explaining
the apparition or desperation of HTML browser quirks.
For browser fingerprinting, this is again very valuable, because it enables us 
to also discriminate between two close browser versions.
For instance, as shown in Table \ref{tab:distancetable2}, we are to very clearly discriminate between  
Safari 4.0.4 and Safari 4.0.5 (distance of 13 much higher than zero).

To conclude, it does not seem possible  to relate the quirks
discrimination power to general factors, while it seems that a
potential explanation may be flaws in the development processes. It is
interesting to observe (see annexes) that the plots are completely
different from one browser family to another.

The classification of web browsers according to quirks must thus
follow another explanation than time. We develop this point in the next section.

\subsection{On Kinds of XSS}
\label{sec:xss-kind}
In this section, we provide some explanations on the discrimination power of HTML parser quirks.
The arguments come from observations done during the experiments, 
as well as from the experience of two authors (junior and senior security engineers in an IT security company).
These arguments form a kind of taxonomy of XSS vectors.
 
\paragraph{Vendor-dependent Vectors} Some vendors (especially Opera and Microsoft Internet Explorer) ship a 
 large variety of features that are unique. 
 This includes CSS expressions, Visual Basic Script support, CSS vendor prefixes such 
 as \textit{-o-link} and other exclusive and often non-standard features. Gecko-based user 
 agents supported by an installed Java Runtime Engine (JRE) and corresponding browser plugin 
 support a non-standard feature called LiveConnect.
 Those unique vendor features often come with XSS holes (vendor dependent vectors), and are gold for fingerprinting.
 For instance, vector \#$397$ 
 selected by the classifier is known to work only under Firefox family browsers.
 
\paragraph{Feature-dependent Vectors} Some XSS vectors depend on a specific feature (yet not vendor specific). 
 Example are the VML-based JavaScript execution and DOM modification 
 vectors functioning in older versions of Microsoft Internet Explorer (IE). Indeed, IE browser is the 
 one supporting the legacy VML feature (a vector graphics format predecessor of SVG -- Scalable Vector Graphics). 
 It has to be noted as that support for this feature started with version 5.5 and ended 
 with version 8. Following versions 9 and 10 are not able to render VML-based images without 
 further effort, document mode switches or additionally loaded behavior files. On the other hand, 
 early versions of Internet Explorer are not capable of displaying SVG images properly -- 
 while IE9 and IE10 do.
 
\paragraph{Version-dependent Vectors} 
  Some quirks are really dependent on the version, especially HTML5-based XSS vectors.
  Partial feature support can usually be detected without large effort 
  and allows very distinct version determination. An example for this classification is the 
  support for features such as Iframe sandboxes and the \textit{srcdoc} functionality. Google 
  Chrome and Webkit browsers implemented partial support for it, and made many minor releases 
   until full its implementation. As a consequence,
  fingerprinting across such minor  versions among the same browser family can be accomplished.   
 
\paragraph{Parser-dependent Vectors} 
Some very discriminant vectors are only dependent on parser specificities such as handling padding 
characters.
Earlier versions of Google Chrome for instance allowed to use 
non-printable characters from the lower ASCII range to be used as padding in URL protocol 
handlers. This strange behavior was later on removed and therefore enables a precise 
fingerprint distinguishing minor versions of Webkit-based browsers. 
Similar effects can be observed when testing against tolerance for white-space and line breaks. 
Man browsers accept exotic characters such as the OGHAM SPACE MARK as valid white space and
therefore semantically relevant part in HTML elements and attributes. Vectors 89,90,128 and 258 
selected by the classifier belong to this category.

\paragraph{Mutation Behavior} Many browsers have slightly different behaviors 
once certain DOM properties are being accessed and mutated: it includes the properties \textit{innerHTML} 
and \textit{cssText}, DOM nodes and CSS objects. Depending on the context and browser version, 
character sequences are being changed, entities are being decoded and escapes removed. 
Special characters and ASCII non-printable may removed or mutated as well -- and
thereby provide yet another goldmine for successful fingerprinting. 
\paragraph{Recapitulation}
There are many sources of HTML parsing specificities (vendors, features, versions, etc.).
The key reason of our fingerprinting capability resides in using all of them in a single unified framework of testable parsing quirks
of the form of XSS vectors.
\begin{table}
  \centering
  \caption{Confusion matrix}
    \begin{tabular}{l|rrrrrr}
    \hline
    classified as & a     & b     & c     & d     & e     & f \\
    \hline
    a = Safari & 11 & 0 & 0 & 0 & 0 & 0 \\
    b = Firefox & 0 & 15 & 0 & 0 & 0 & 0 \\
    c = IE & 0 & 0 & 6 & 0 & 0 & 0 \\
    d = Opera & 0 & 0 & 0 & 6 & 0 & 0 \\
    e = Android & 0 & 0 & 0 & 0 & 14 & 1 \\
    f = Chrome & 0 & 0 & 0 & 0 & 0 & 19 \\
    \hline
    \end{tabular}
  \label{tab:confusionmatrix}
\end{table}

\subsection{Limitations}
\label{sec:limitations}
We now discuss the important limitations of our approach.

First, a  common weakness of browser fingerprinting tools is that responses
from the browser can be forged by the attacker: the proposed technique does not offer 
an exception to this rule. 
To spoof a victim XSS-based fingerprint, an attacker must
 either emulate the behavior of a specific web browser or have an  adaptation environment enabling the deployment of the appropriate web browser at runtime.
The economical aspect of security is a key factor in cyber-attacks, and our technique makes user-agent based attacks more costly for the attacker.

Second, we have shown that our technique enables defenders to precisely determine the browser family and version.
However, in reality, most users will be using IE, Chrome, or Firefox at their latest versions. 
In other terms, an attacker would just have to deploy a handful of browsers at runtime as a counter measure to our fingerprinting approach.
This is not only a limitation, in order to defeat spoofing of our fingerprinting technique,  it is a good idea to use a "rare" browser, both in terms of family and version.

Third, XSS bugs get fixed over time. This may be a limitation since our fingerprinting capabilities may decrease over time. So far, this is not true.
According to our empirical data, until now, the rate of XSS introduction (due to new features) is comparable to the rate of XSS removals (due to bug fixing).

Finally, the technique we propose only considers the quirks related to html parsing, which can be seen as a limitation. Our technique cannot fully protect a defender but 
should  be used as a lightweight technique to be used in complement to more heavy-weight techniques (see related work).

\section{Other Uses of Browser Fingerprinting}
\label{sec:other-uses}
Whatever the fight is, when the weapons are comparable, harming a target 
requires the identification of weaknesses to adapt the attack accordingly. 
Conversely, defending from an attacker also requires a similar analysis that 
enables an appropriate counter-attack. Besides, both opponents will develop 
their own protecting measures, improving the armor they wear; history has 
shown many examples of such up-to-extreme improvements (e.g. plate armors 
of late occidental Middle Age).

This symmetrical aspect of a fight, with the same offensive weapons, also occurs
in nowadays web security, in which the notion of counter-attack is becoming
crucial. While an attacker will try to identify the exact web browser his
victim uses to imagine a dedicated attack, a defender of a web site may want
to detect the exact web browser the attacker uses, improving his ability of
counter-attacking him.

In this section, we describe such sophisticated couter-measures and malicious usage 
of browser fingerprinting
from the viewpoint of both security engineers and malicious attackers.

\subsection{Browser Exploit Kits}
Malware propagation via browsers is done through browsers exploit kits. This is 
a piece of server side software that fingerprints client browsers in order to 
deliver malware. 
Users are attracted to such malicious servers through advertisement systems or 
compromised websites. 
For instance, users presenting a Firefox user agent receive an specific 
exploit based on its version.
These exploits are written in JavaScript for browser exploits, or in plugin 
specific languages (VBScript, ActiveX, Java, Flash \ldots) for plugin specific 
exploits.

Browser exploit kits mainly use User Agent to naively fingerprinting 
browsers.
Browser exploit kits rely on browser specific capabilities (DOM Tree, VBScript
execution \ldots).
At the time of writing, only specific JavaScript engine behaviors 
\cite{egele2009defending} are used as an advanced browser fingerprinting 
mechanism, but very few studies are available on the subject.
Browser exploit kits will implement more and more advanced browser fingerprinting 
mechanisms. Studying them improve our understanding of these future issues for malware fighters.

\subsection{Defense Using Client Side Honeypots}
A client side honeypot is a browser like application suited to collect 
browser exploits and malware samples when visiting an website suspected to host 
a browser exploit kit \cite{provos2004virtual}. Two family of honeypot exists, 
low interaction, and high
interaction honeypot clients (or honey-clients).

Low interaction ones like \textit{honeyc} \cite{seifert2007honeyc} are made of 
spoofed browser User Agent and just follow links provided by exploit kits and
collects any executable they find. These pieces of malware are then 
automatically submitted to malware analysis platforms like Anubis 
\cite{bayer2006dynamic}. By spoofing various popular user agents and iterating
connections on exploit kit URL, a single honey-client can collect a subsequent
amount of browser exploits. 
However, if the browser exploit kit uses advanced browser fingerprinting, such
low interactions honey-client fail to identify malicious website and to collect
malware.

To overcome this problem, high interaction honey-clients combine are made of 
instrumented browsers running into virtual machines like \textit{phoneyc} 
\cite{nazario2009phoneyc}. 
``High-interaction'' means that the honey-client can respond to all kind of 
fingerprinting challenges sent by the browser exploit kit (such as JavaScript 
execution).
This approach is very heavyweight.
By knowing browser fingerprints summarizing high interaction 
fingerprinting challenges, low interaction client side honeypots are much 
easier to build and maintain compared to high interaction honey-clients.

\subsection{Detection of XSS Proxification}
XSS proxification consists of using a cross-site scripting (XSS) vulnerability on a website to 
force the victim's browser to request web pages on behalf of an attacker and to send the result 
back to it. In other words, it turns the victim browser in a traditional HTTP Proxy. 
The beef project tunneling proxy features implement such an 
attack\footnote{\url{https://github.com/beefproject/beef/wiki/Tunneling-Proxy}}. 
Detection of XSS proxification with all kinds of techniques based on TCP network shape, 
HTTP headers (incl. user-agent) and IP addresses is vain, since the infected browser itself does 
the request.
However, browser fingerprinting can be used to detect XSS proxification since the 
browser engine of the attacker is likely to be different from the infected engine.

\subsection{Detection of Disguised Crawlers}
Malicious crawlers tend to use user-agents strings of standard client browsers.
On the one hand, they don't have to declare themselves, on the other hand, this allows them 
to access resources that are restricted to robots and crawlers.
Detecting disguised crawlers is especially important to ban clients that are eating all 
resources up to all kinds of deny-of-service.
We think that techniques based on browser fingerprinting may be used to detect whether a client is a 
bot or not.
\section{Related Work}
\label{sec:related}

\subsection{Passive OS Fingerprinting (pOf)}
\label{subsec:pof}
In this paper, Lippmann et al. show how OS fingerprinting could be a major advantage in 
Intrusion Detection Systems: the use of the surface of attack of an OS permits to dismiss 
an alert when a vulnerability cannot be exploited for the identified 
OS~\cite{lippmann2003passive}. We can exploit our works in the same way. The objectives of the 
paper are to demonstrate 
1) how pOf is used to determine accurately OS by analyzing TCP/IP packet 
2) the evaluation of pOf tools and
3) the assessment of a new classifier using data mining and pattern classification techniques.
The main difference between pOf and active fingerprinting is that pOf does not send frames to 
the targeted host but instead analyses headers of packets exchanged during normal traffic. 
Thus, pOf is less accurate than active OS fingerprinting. Different classifier techniques are 
presented and evaluated: \textit{k-nearest neighbor} (KNN), 
binary tree, Multi-Layer Perceptrons (MLP) and Support Vector Machine (SVM). 
The confidence in a technique depends on the number of fields analyzed during pOf.

\subsection{Passive Fingerprinting of User Agent from Network Flow Logs}
\label{subsec:passive-fp-network-flow}
    
Yen et al. use machine learning to passively fingerprint browsers based on their network 
behavior~\cite{yen2009browser}. The number of TCP connections launched, number of requests 
and frequency, all these parameters are dependent of the browser implementation and provide a 
Fingerprint that can be automatically built out of Bayesian belief networks. The main advantage 
of this technique is that it only needs coarse traffic summaries to identify the browser family.  
They use two techniques to classify browser: per-browser or generic classifiers with a maximum 
difference in precision of 15\%.
\textit{Our technique is more accurate since it can fingerprint browser versions}
    
\subsection{Fingerprinting using Browser Scripting Environment}
    
Fioravanti proposes usage of various JavaScript features and specific API elements to determine 
the browser family~\cite{fioravanti2010client}. But these elements collected from JavaScript 
can be altered by the usage of a specific plugin (like user-agent switcher in Firefox) or by 
overwriting the tests results with the correct values. 
\textit{The main difference of our approach is that it uses HTML parser specificities, much harder
to spoof because it requires the same database of quirks than the fingerprinting database,
and modification in the parsing engine itself to implement the behavior.}

\subsection{Panopticlick: Browser Uniqueness Fingerprint}
\label{subsec:panopticlick}
In this paper, Eckersley et al. collect bits of information from various browser properties 
(user agent string, screen resolution, installed fonts and plug-ins) to fingerprint the user 
browser~\cite{eckersley2010unique}. These pieces of information are collected through Java, Flash, and
JavaScript. Using all these properties a user can sometimes be uniquely identified. 
Compared to our work, the differences are important.
First, uniquely identifying a browser instance does not necessarily imply knowing the browser type and version for attacks or counter-measures.
Second, Panopticlick uses Java, Flash, and JavaScript, which is a stronger assumption on the client browsers than ours (we only rely on HTML).
However, we think that it would be an interesting area of future work to combine our approach with Java, Flash, or JavaScript fingerprinting mechanisms.

\subsection{Fingerprinting Information in JavaScript Implementation}
    
Mowery et al. use measures from 39 performance tests to generate a signature in the form of a 39 
dimension vector representing test timing results~\cite{mowery2011fingerprinting}. They have a 
browser family detection rate of 98.2\% in the conditions of the experiment. But when dealing with subversions of given browsers, the precision drops to 79.8\% for major version identification. The most interesting contribution is the underlying architecture fingerprinting capability.
\section{Conclusion}
\label{sec:conclusion}
In this paper, we have presented an approach to fingerprinting web browsers based on XSS vectors.
This approach is able to perfectly fingerprint 78\% of our browser dataset.
To fingerprint only the browser family, the recognition ratio is 98\% with only six XSS vectors to be executed.
We are now working on extending our browser signature database using Amazon's Mechanical Turk. 
We also plan to mix different browser fingerprinting techniques (JavaScript, network traffic, etc.) to achieve even higher recognition rates.
\bibliographystyle{IEEEtran} 
\bibliography{u4fa3ca161ba56}

\begin{table*}[h]
  \centering
  \caption{Distance analysis using Modified Hamming Distance  (first part)}
    \begin{tabular}{llrrrr}
    \hline
    Browser & Nearest Neighbor (MHD) & MHD   & MDF   & Fsize & MDD \\
    \hline
    \#89 - Origin Browser & \#28 - Safari 5.1.5/MacOSX 10.7.3 & 3     & -     & 1     & 129.0 \\
    \#25 - fbx v6 & \#8 - Safari 5.1.5 & 7     & -     & 1     & 127.5 \\
    \#27 - Rekonq Linux & \#40 - Safari 5.0.6 & 15    & -     & 1     & 131.0 \\
    \#11 - Konqueror 4.7.4/KHTML & \#46 - Chrome 3.0.182.2 & 52    & -     & 1     & 88.5 \\
    \#5 - Firefox 11.0/Win7 & \#39 - Firefox 11.0 & 0     & 0,5   & 15    & 67.5 \\
    \#9 - Firefox 10.0/Ubuntu/Linaro & \#39 - Firefox 11.0 & 0     & 0,5   & 15    & 67.5 \\
    \#16 - Mozilla Firefox 11.0  Ubuntu & \#39 - Firefox 11.0 & 0     & 0,5   & 15    & 67.5 \\
    \#21 - Firefox 10 & \#39 - Firefox 11.0 & 0     & 0,5   & 15    & 62.5 \\
    \#39 - Firefox 11.0 & \#59 - Mozilla Firefox 9.0 & 0     & 0,5   & 15    & 67.5 \\
    \#51 - Mozilla Firefox 8.0 & \#39 - Firefox 11.0 & 0     & 0,5   & 15    & 67.5 \\
    \#59 - Mozilla Firefox 9.0 & \#39 - Firefox 11.0 & 0     & 0,5   & 15    & 67.5 \\
    \#60 - Mozilla Firefox 10.0 & \#39 - Firefox 11.0 & 0     & 0,5   & 15    & 67.5 \\
    \#4 - Firefox 8.0.1 & \#88 - Firefox 11.0 linux & 0     & 1     & 15    & 68.5 \\
    \#88 - Firefox 11.0 linux & \#4 - Firefox 8.0.1 & 0     & 1     & 15    & 68.5 \\
    \#62 - Chrome 12.0.742.91 & \#63 - Chrome 13.0.782.99  & 0     & 2     & 19    & 71.5 \\
    \#63 - Chrome 13.0.782.99  & \#62 - Chrome 12.0.742.91 & 0     & 2     & 19    & 71.5 \\
    \#58 - Chrome 10.0.648.133 & \#57 - Chrome 9.0.597.94 & 1     & 3     & 19    & 72.5 \\
    \#1 - Chrome 18.0 & \#15 - Chromium 18.0 & 0     & 3,5   & 19    & 69.5 \\
    \#15 - Chromium 18.0 & \#65 - Chrome 16 & 0     & 3,5   & 19    & 69.5 \\
    \#64 - Chrome 14.0.814.0 & \#15 - Chromium 18.0 & 0     & 3,5   & 19    & 69.5 \\
    \#65 - Chrome 16 & \#15 - Chromium 18.0 & 0     & 3,5   & 19    & 69.5 \\
    \#70 - Chrome 17.0.963.8 & \#15 - Chromium 18.0 & 0     & 3,5   & 19    & 69.5 \\
    \#75 - Chrome 18 / Win XP 32 & \#15 - Chromium 18.0 & 0     & 3,5   & 19    & 69.5 \\
    \#66 - Chrome 15.0.874.106 & \#15 - Chromium 18.0 & 1     & 3,5   & 19    & 69.5 \\
    \#56 - Chrome 8.0.552.215 & \#57 - Chrome 9.0.597.94 & 0     & 4     & 19    & 73.5 \\
    \#57 - Chrome 9.0.597.94 & \#56 - Chrome 8.0.552.215 & 0     & 4     & 19    & 73.5 \\
    \#83 - Firefox 11.0 & \#4 - Firefox 8.0.1 & 4     & 5     & 15    & 70.0 \\
    \#19 - Firefox 7.0 & \#39 - Firefox 11.0 & 5     & 5     & 15    & 70.5 \\
    \#55 - Chrome 7.0.517.41 & \#57 - Chrome 9.0.597.94 & 3     & 7     & 19    & 72.5 \\
    \#53 - Chrome 6.0.453.1 & \#57 - Chrome 9.0.597.94 & 7     & 7,5   & 19    & 72.0 \\
    \#96 - Chrome Nexus S & \#15 - Chromium 18.0 & 6     & 8,5   & 19    & 69.5 \\
    \#73 - Chrome 18.0 & \#15 - Chromium 18.0 & 9     & 11    & 19    & 76.5 \\
    \#68 - Opera 11.65 Mac OS X 10.7.3 & \#2 - Opera 11.11 & 9     & 14    & 6     & 124.0 \\
    \#107 - IE 9 & \#3 - IE 9.0 & 9     & 17,5  & 6     & 69.0 \\
    \#24 - IE 7.0 & \#86 - IE 7.0 & 1     & 21    & 6     & 76.0 \\
    \#86 - IE 7.0 & \#24 - IE 7.0 & 1     & 21    & 6     & 77.0 \\
    \#2 - Opera 11.11 & \#7 - Opera 11.52/Win7 & 3     & 21    & 6     & 136.0 \\
    \#84 - IE 7.0 & \#86 - IE 7.0 & 4     & 22    & 6     & 78.0 \\
    \hline
    \end{tabular}
  \label{tab:distancetable1}
\end{table*}
\begin{table*}[h]
  \centering
  \caption{Distance analysis using Modified Hamming Distance (second part)}
    \begin{tabular}{llrrrr}
    \hline
    Browser & nearest neighbor (MHD) & MHD   & MDF   & Fsize & MDD \\
    \hline
    \#7 - Opera 11.52/Win7 & \#2 - Opera 11.11 & 3     & 24    & 6     & 134.0 \\
    \#18 - Opera 11.62 & \#68 - Opera 11.65 Mac OS X 10.7.3 & 14    & 24    & 6     & 133.0 \\
    \#31 - Firefox 3.0.17 & \#32 - Firefox 3.0.15 & 0     & 25    & 15    & 79.5 \\
    \#32 - Firefox 3.0.15 & \#31 - Firefox 3.0.17 & 0     & 25    & 15    & 79.5 \\
    \#29 - Firefox 3.0.6 & \#31 - Firefox 3.0.17 & 2     & 25    & 15    & 81.5 \\
    \#85 - IE 8.0 & \#107 - IE 9 & 23    & 25,5  & 6     & 100.0 \\
    \#95 - Android 2.3.3 & \#94 - ANdroid 2.3.1 & 13    & 26    & 15    & 160.5 \\
    \#100 - Samsung galaxy ace & \#105 - Samsung Galaxy S & 13    & 26,5  & 15    & 151.0 \\
    \#104 - LG p970 & \#106 - Sony Xperia s & 11    & 27    & 15    & 142.0 \\
    \#94 - ANdroid 2.3.1 & \#95 - Android 2.3.3 & 13    & 27    & 15    & 154.5 \\
    \#106 - Sony Xperia s & \#104 - LG p970 & 11    & 27,5  & 15    & 152.5 \\
    \#101 - Samsung galaxy y & \#100 - Samsung Galaxy Ace & 13    & 29    & 15    & 154.5 \\
    \#105 - Samsung galaxy s & \#100 - Samsung Galaxy Ace & 13    & 30    & 15    & 155.0 \\
    \#48 - Chrome 4.0.223.11 & \#52 - Chrome 5.0.307.1 & 4     & 31    & 19    & 73.5 \\
    \#52 - Chrome 5.0.307.1 & \#48 - Chrome 4.0.223.11 & 4     & 31    & 19    & 75.5 \\
    \#98 - Samsung galaxy tab & \#104 - lg p970 & 15    & 31    & 15    & 157.0 \\
    \#3 - IE 9.0 & \#107 - IE 9 & 9     & 32,5  & 6     & 85.0 \\
    \#17 - Internet Explorer 9 Win 7 64b & \#107 - IE 9 & 15    & 35    & 6     & 80.0 \\
    \#46 - Chrome 3.0.182.2 & \#48 - Chrome 4.0.223.11 & 10    & 37    & 19    & 64.5 \\
    \#6 - Opera 12/Android 2.3.3 & \#68 - Opera 11.65 Mac OS X 10.7.3 & 27    & 37    & 6     & 127.0 \\
    \#79 - Android 1.5 & \#80 - Android 1.6 & 19    & 39    & 15    & 144.0 \\
    \#80 - Android 1.6 & \#79 - Android 1.5 & 19    & 41,5  & 15    & 147.0 \\
    \#99 - HTC Desire hd & \#100 - Samsung Galaxy Ace & 40    & 44,5  & 15    & 151.5 \\
    \#82 - Android 2.1 & \#95 - Android 2.3.3 & 38    & 47    & 15    & 158.5 \\
    \#37 - Opera 10.6 & \#2 - Opera 11.11 & 41    & 49    & 6     & 127.0 \\
    \#92 - Safari 3.2.1 & \#91 - Safari 3.1.2 & 1     & 54    & 11    & 149.0 \\
    \#91 - Safari 3.1.2 & \#92 - Safari 3.2.1 & 1     & 56,5  & 11    & 148.0 \\
    \#69 - Safari 4.0.4 & \#90 - Safari 4.0.5 & 13    & 60,5  & 11    & 148.5 \\
    \#81 - Safari 5.0.5 & \#69 - Safari 4.0.4 & 20    & 64,5  & 11    & 152.5 \\
    \#40 - Safari 5.0.6 & \#8 - Safari 5.1.5 & 9     & 65    & 11    & 126.0 \\
    \#90 - Safari 4.0.5 & \#69 - Safari 4.0.4 & 13    & 65,5  & 11    & 157.0 \\
    \#28 - Safari 5.1.5/MacOSX 10.7.3 & \#89 - Origin Browser & 3     & 68    & 11    & 132.0 \\
    \#8 - Safari 5.1.5 & \#25 - fbx v6 & 7     & 68,5  & 11    & 121.5 \\
    \#87 - Safari iPhone & \#40 - Safari 5.0.6 & 25    & 74    & 11    & 138.0 \\
    \#23 - Safari 5 Windows 7 64b & \#8 - Safari 5.1.5 & 19    & 75    & 11    & 119.5 \\
    \#103 - Android 3.0 & \#28 - Safari 5.1.5/MacOSX 10.7.3 & 21    & 78    & 15    & 135.0 \\
    \#93 - Safari 3.0.4 & \#92 - Safari 3.2.1 & 41    & 81,5  & 11    & 181.0 \\
    \#74 - Samsung GT-S5570 Android & \#11 - Konqueror 4.7.4/KHTML & 116   & 139   & 15    & 137.5 \\
    \#97 - Google Samsung Nexus & \#96 - Chrome Nexus S & 10    & 151,5 & 15    & 69.5 \\
    \hline
    \end{tabular}
  \label{tab:distancetable2}
\end{table*}

\end{document}